\documentclass{ws-ijqi}
\usepackage{bbm}
\begin{document}
\markboth{A. Mandarino et al}
{About the use of fidelity in CV systems}
\catchline{}{}{}{}{}
\title{ABOUT THE USE OF FIDELITY \\ IN CONTINUOUS VARIABLE SYSTEMS}
\author{ANTONIO MANDARINO, MATTEO BINA}
\address{Dipartimento di Fisica dell'Universit\`a degli 
Studi di Milano 20133 Milano, Italy\\
antonio.mandarino@unimi.it}
\author{STEFANO OLIVARES, MATTEO G.~A.~PARIS}
\address{Dipartimento di Fisica, Universit\`a degli Studi di Milano, 
I-20133 Milano, Italy\\
CNISM, UdR Milano, I-20133 Milano, Italy
}\maketitle
\begin{history}
\received{\today}
\end{history}
\begin{abstract}
We present examples of continuous variable (CV) states having high fidelity
to a given target, say $F>0.9$ or $F>0.99$, and still showing striking
differences in their physical properties, including classical and
quantum states within the set, separable and entangled ones, or
nearly Gaussian and strongly non-Gaussian ones.
We also show that the phenomenon persists also when one imposes additional
constraints on the energy or the squeezing fraction of the states, thus
generally questioning the use of fidelity to assess properties 
of CV systems.
\end{abstract}
\section{Introduction}	
Fidelity~\cite{Uhl}  is a widely adopted figure of merit to compare 
quantum states and to assess generation and characterization schemes 
of interest for quantum technology, e.g. quantum interferometry
\cite{cav81,par95,ban09}. 
Fidelity between the two states $\rho_1$
and $\rho_2$ is defined as follows:
\begin{equation}
F(\rho_1,\rho_2) = \left( \mbox{Tr} \sqrt{\sqrt{\rho_1} 
\rho_2 \sqrt{\rho_1} }  \right)^2\,.
\end{equation}
The Bures distance may be expressed in terms of fidelity:
\begin{equation}
D_B(\rho_1,\rho_2)=\sqrt{2 \left[ 1-\sqrt{F(\rho_1,\rho_2)} \right]}\,,
\end{equation}
which also provides an upper and lower bounds to the trace 
distance:~\cite{fuc99}
\begin{equation}
1-\sqrt{F(\rho_1,\rho_2)}\leq \frac12 || \rho_1-\rho_2||_1\leq
\sqrt{1-F(\rho_1,\rho_2)}\,.
\end{equation}
\par
Fidelity is bounded to the interval $[0, 1]$, and values 
above a given threshold close to unit, say, 0.9 or 0.99 are usually considered 
as a sign that the two states are close to each other, and so share nearly 
identical properties.  The first statement is certainly true, as it follows 
from the links between the fidelity and the Bures and trace 
distances, whereas the second one may be not justified, or even wrong 
in some cases.~\cite{edvd,dodonov} The main purpose of this paper is to 
continue and extend the analysis of Ref.~\refcite{Bina}, providing 
examples, in CV systems, where high values of fidelity 
are achieved by pair of states with considerably different physical 
properties, as for example separable and entangled states, classical 
and nonclassical ones or, going beyond the Gaussian sector, states 
with very different values of non-Gaussianity.
\par
The paper is structured as follows. In Section~\ref{s:1} we deal with
single-mode Gaussian states and analyze the drawbacks of the use of
fidelity is assessing their quantumness, defined either in terms of
Glauber $P$-functions or via the Fano factor. In Section~\ref{s:2} we
focus on two-mode states and show how high values of fidelity may be
achieved by separable and entangled states or by states with very
different values of non-Gaussianity.  Section~\ref{s:out} closes the
paper with some concluding remarks.
\section{Single-mode Gaussian State}
\label{s:1}
Here we consider a generic single-mode Gaussian state that is a
displaced squeezed thermal state (DSTS$_1$) :
\begin{equation} {\label{SMGaussian}}
\rho(x, r, n_T) = D(x)S(r)\nu_{\text{th}}(n_T)S^\dag(r)D^\dag(x) \,,
\end{equation} 
where $S(r)=\exp \{ \frac{1}{2} r (\hat{a}^{\dag2}-\hat{a}^2 )\} $ and
$D(x)=\exp \{ x( \hat{a}^\dag-\hat{a}) \} $, with $ r,x \in \mathbbm{R}$, are  the
single-mode squeezing and the displacement operators, respectively,
$\nu_{\text{th}}(n_T)$ is a thermal state with 
mean photon number $n_T$.
The covariance matrix (CM) of the state in (\ref{SMGaussian}) 
is diagonal $\mathbf{\sigma}=\hbox{diag} (a,b)$
with $a=(n_T+ \frac{1}{2})\, e^{2r}$, $b=(n_T+ \frac{1}{2})\, e^{-2r}$.
A suitable parametrization for DSTS$_1$ may be obtained using the
coherent amplitude $x$, the average photon number of the squeezed
thermal kernel $\rho(0, r, n_T)$, i.e. $N= n_T + n_S + 2  n_T n_S$
where $n_T$ is thermal photon numbers and 
$n_S= \sinh^2 r$ is the squeezing contribution, and the
squeezing fraction $\beta\equiv n_s/N \in [0,1]$.
The total average photon number of DSTS$_1$ (from now on the {\em
energy}) is given by $\langle a^\dag a \rangle \equiv \langle n \rangle =
x^2 +N$ and the thermal and squeezing component may be expressed as:
\begin{equation}
n_T=  \frac{(1- \beta)N}{1 + 2\beta N}  \qquad \text{and} \qquad n_S=\beta N\,,
\end{equation}
respectively.
\par
\begin{figure}[h!]
\center
\begin{minipage}{0.49\columnwidth}
\centerline{
\includegraphics[width=0.95\columnwidth]{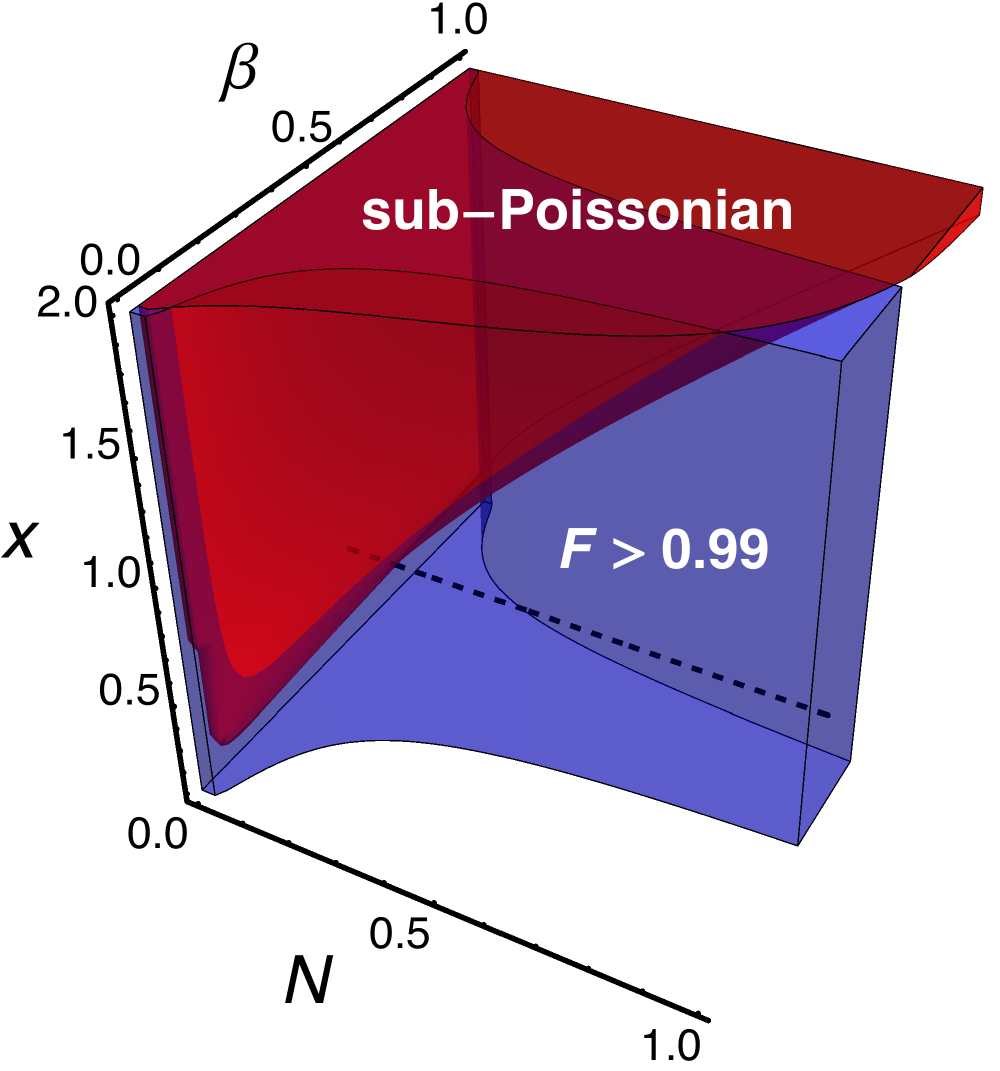}
}
\end{minipage}
\begin{minipage}{0.49\columnwidth}
\centerline{
\includegraphics[width=0.95\columnwidth]{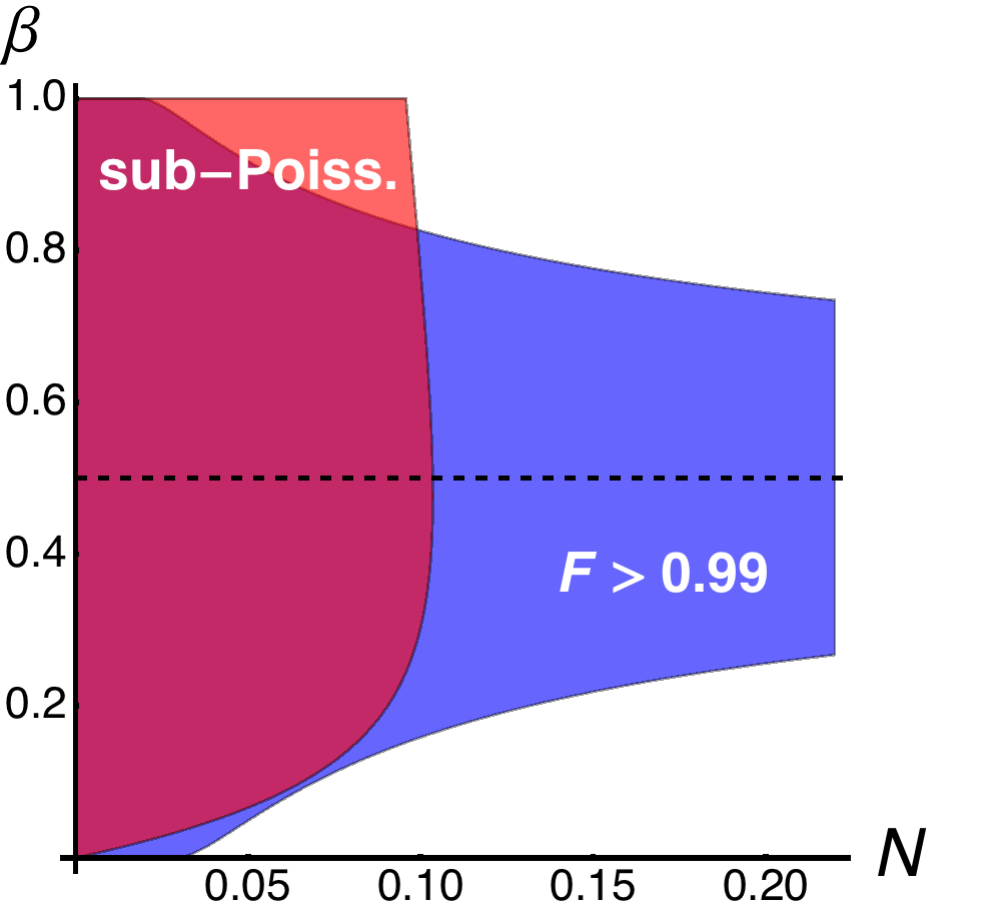}
}
\end{minipage}
\caption{\label{fani} (Color online) Fidelity and sub-Poissonianity.
(Left panel) The red region contains the sub-Poissonian DSTS$_1$ whereas
the blue one refers to states with fidelity $F_{N \beta x}>0.99$ to a target DSTS$_1$
with the same $N$  and fixed $\beta=0.5$ and $x=0.5$ (black-dashed line).
(Right panel) Section of the plot of the left panel in correspondence of $x=0.5$.
Note that here $N$ is not the total energy, but the average photon
number of the squeezed thermal kernel (see text for details).} 
\end{figure}

The nonclassicality of a DSTS$_1$ may be detected using the Fano factor
defined as the ratio of the variance of photon number over the mean
photon number:~\cite{HP82}
\begin{equation} \label{FanoFac}
R=
\frac{\langle n^2 \rangle -\langle n \rangle ^2}{\langle n \rangle}\,.
\end{equation}
One has $R =1$ for coherent states, while a smaller value
is a sufficient condition for  nonclassicality, since no state endowed with a positive
Glauber $P-$function may be sub-Poissonian. The fidelity between two
single-mode Gaussian states $\rho_k(x_k,r_k,n_{T,k})$ and $k=1,2$
may be written as:~\cite{twa96,scu98}
\begin{equation}
F_{N \beta x } = \frac{\exp\{ ( \bf{X_1}-\bf{X_2})^T (\sigma_1+\sigma_2)^{-1}
( \bf{X_1}-\bf{X_2})\}}{\sqrt{\Delta + \delta}-\sqrt{\delta}} \,,
\end{equation}
where $\sigma_1$ and $\sigma_2$ are the corresponding covariance
matrices,  $\Delta=\det[\sigma_1 + \sigma_2]$, $\delta=4\prod_{k=1}^2 
(\det[\sigma_k]-\frac{1}{4})$, and where $\bf{X_k}$=$(x_k,0)$.
In Fig.~\ref{fani} (left panel) we show DSTS$_1$ as points in the space 
parametrized by $N$, $\beta$ and $x$: the red region 
corresponds to sub-Poissonian states, whereas the blue one 
contains states having fidelity $F_{N \beta x} > 0.99$ to a DSTS$_1$
target with the same value of $N$, that is the average photon
number of the squeezed thermal kernel, and  $\beta=0.5$ and 
$x=0.5$ (dashed line in the figure). For the sake of clarity, we report in 
the right panel of  Fig.~\ref{fani} a section of the left panel for $x=0.5$.
As it is apparent from the plot, this set includes both 
sub-Poissonian and super-Poissonian states, independently on the 
nature of the target state. Overall, this means that fidelity cannot 
be used to assess the sub-Poissonian character
of DSTS$_1$ even when quite strict constraints are imposed 
on the set of considered states.
\par
\begin{figure}[h!]
\center
\begin{minipage}{0.49\columnwidth}
\centerline{
\includegraphics[width=0.9\columnwidth]{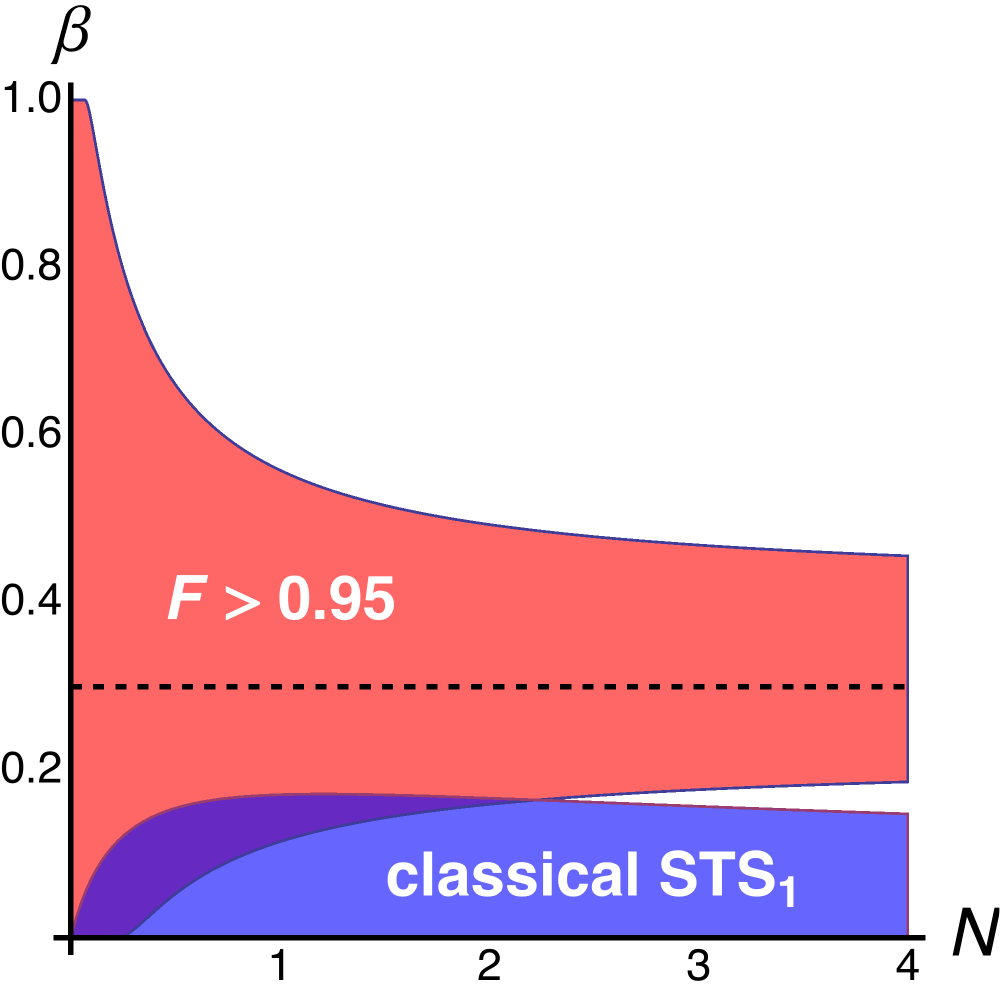}
}
\end{minipage}
\begin{minipage}{0.49\columnwidth}
\centerline{
\includegraphics[width=0.9\columnwidth]{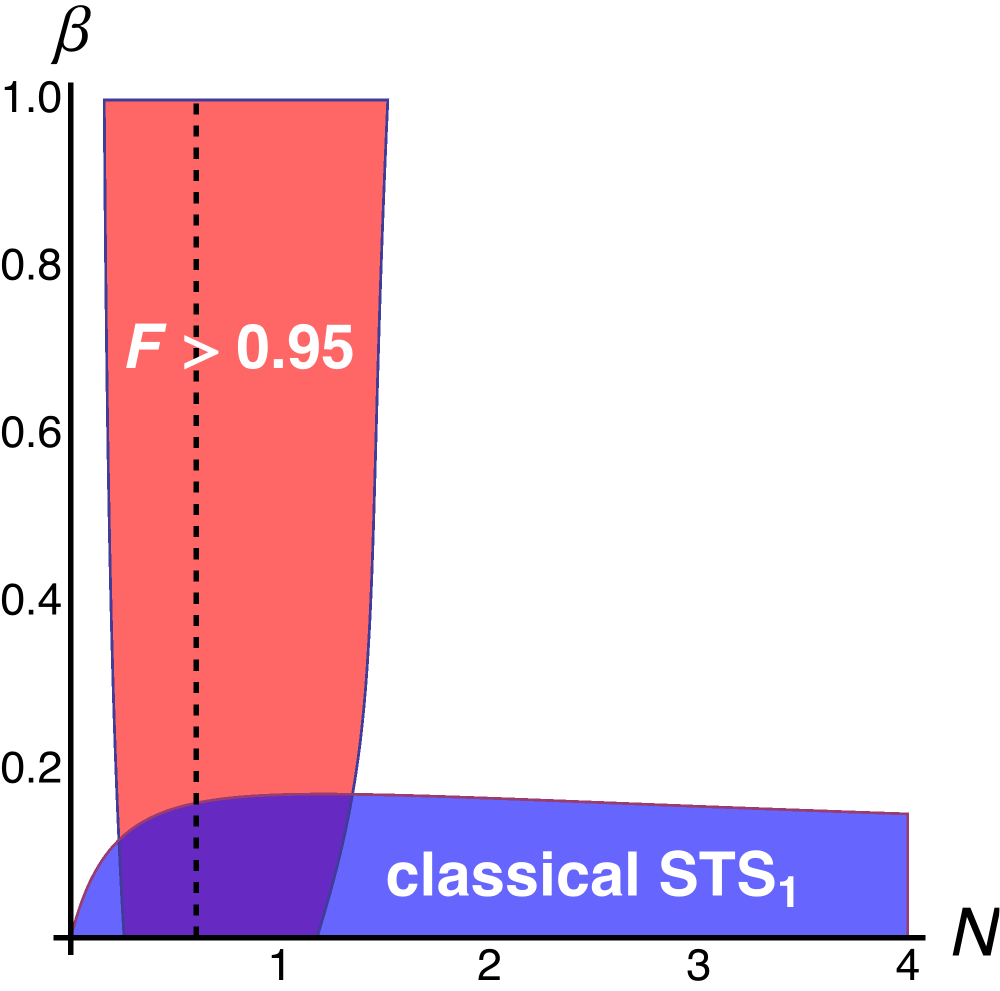}
}
\end{minipage}
\caption{\label{noncl} (Color online) Fidelity and nonclassicality.
Classicality regions for STS$_1$ (blue regions) and regions of
states having fidelity $F_{N \beta}> 0.95$ (red regions)
to the set of target states (black-dashed line) as functions of $N$
and $\beta$. The target states have fixed $\beta=0.3$ (left panel) or
fixed $N=0.6$ (right panel).}
\end{figure}

The most general way to assess the quantum properties of a 
single-mode state is to study whether the Glauber $P-$function is singular
or not. Let us focus attention  to single-mode squeezed thermal states,
i.e. let us set $x=0$ in Eq.~(\ref{SMGaussian}), and analyze the relationships
between nonclassicality and fidelity.
In the left panel of Fig.~\ref{noncl} we show the region of 
classicality for STS$_1$ states as a function of the total energy $\langle n\rangle =N$ 
and the squeezing fraction $\beta$ together with the region of states having a fidelity 
$F_{N \beta}> 0.95$ to the set of nonclassical states having 
fixed squeezing fration $\beta=0.3$.
The right panel of Fig.~\ref{noncl} displays the region of 
classical states together with the region of states
having a fidelity $F_{N \beta}> 0.95$ to the set of states 
with fixed energy $N=0.6$. In both cases the areas have a nonzero
overlap and cross the non classical boundary, such that fidelity cannot 
be used as unique figure of merit in order to assess quantumness. 
To summarize: we have strong evidence that fidelity should not be used 
in benchmarking the generation  of quantum resources, even when
attention is focused on states with quite stringent physical 
constraints, as fixed energy or squeezing. Only after a full tomographic 
reconstruction of the state one obtains a suitable set of physical 
quantities to properly decrease the area of
accessible states, having a given value of fidelity to the 
target or the set of target states.~\cite{jar09} 
\section{Two-mode states}
\label{s:2}
Let us now consider two-mode squeezed thermal states, expressed
by the density operator:
\begin{equation}{\label{2modeGaussian}}
\rho = S_2(r)\nu_{\text{th}}(n_{T1})\otimes\nu_{\text{th}}(n_{T2})S_2^\dag(r)\,,
\end{equation} 
where $S_2(r)=\exp\{r(\hat{a}^\dag  \hat{b}^\dag-\hat{a}\hat{b})\}$, with $r\in\mathbb{R}$, is
the two-mode squeezing operator and $n_{Tk}$ ($k=1,2$) are the mean
thermal photon numbers. 
The states in Eq.~(\ref{2modeGaussian}) are Gaussian, assuming 
 their $4 \times 4$ CM is given by:
\begin{equation}
{\mathbf{\sigma}}=\frac{1}{2}\begin{pmatrix}
A\, \mathbb{I}_2& C\, \hat{\sigma}_z\\
C\, \hat{\sigma}_z & B\, \mathbb{I}_2
\end{pmatrix}\,,
\end{equation}
where $ \mathbb{I}_2$ is the $2\times 2$ identity matrix,
$\hat{\sigma}_z$ is the Pauli matrix and:
\begin{subequations}\begin{align}
A \equiv A(\beta,\gamma,N)&=1+\frac{2\gamma(1-\beta)N+\beta N (1+N)}{1+\beta N}\,, \\
B \equiv B(\beta,\gamma,N)&=1+\frac{2(1-\gamma)(1-\beta)N+\beta N (1+N)}{1+\beta N}\,, \\
C \equiv C(\beta,\gamma,N)&=\frac{(1+N)\sqrt{\beta N(2+\beta N)}}{1+\beta N}\,,
\end{align}\end{subequations}
where now 
$N\equiv2n_s+(n_{T1}+n_{T2})(1+2n_s)$
is the total energy, $\beta\equiv 2n_s/N$ is the
fraction of squeezed photons, and $\gamma\equiv n_{T1}/(n_{T1}+n_{T2})$
is the fraction of single-mode thermal photons.
\par
A way to quantify non-classical features of a Gaussian state is the amount of
entanglement. The entanglement could be attested in terms of positivity
 of the partial transposed density matrix. Since the corresponding
 symplectic eigenvalues are:~\cite{Oli12}
\begin{equation}\label{ppt}
\tilde{d}_\pm=
\sqrt{\frac{\tilde{\Delta}(\sigma)\pm\sqrt{\tilde{\Delta}(\sigma)^2-4I_4}}{2}},
\end{equation}
where  we introduced the local symplectic invariants
$I_1= \det[A]$, $I_2= \det[B]$, $I_3= \det[C]$ and $I_4= \det[\sigma]$, 
a two-mode squeezed thermal state is
separable iff $\tilde{d}_- \geq \frac{1}{2}$.  The fidelity between
two-mode Gaussian states of the type (\ref{2modeGaussian}) reads:~\cite{sorin,Marian}
\begin{equation}
F_{N \beta \gamma}= \frac{(\sqrt{{X}}+\sqrt{{X}-1})^2}{\sqrt{\det[\sigma_1+\sigma_2]}}
\end{equation}
where 
$X=2\sqrt{{E_1}}+2\sqrt{{E_2}}+\frac{1}{2}$ and:
\begin{equation}
E_1=\frac{\det[\Omega \sigma_1 \Omega \sigma_2]-
\frac14}{\det[\sigma_1+\sigma_2]}\qquad \mbox{and} \qquad E_2 = 
\frac{ \det[\sigma_1 + \frac{i}{2} 
\Omega]\det[\sigma_2 + \frac{i}{2} \Omega]}{\det[\sigma_1+\sigma_2]},
\end{equation}
$\Omega= i\, \hat{\sigma}_y \oplus \hat{\sigma}_y$ being the $2$-mode symplectic form where
$\hat{\sigma}_y$ is the Pauli matrix.
\par
\begin{figure}[h!]
\centering
\begin{minipage}{0.49\columnwidth}
\centerline{
\includegraphics[width=0.9\columnwidth]{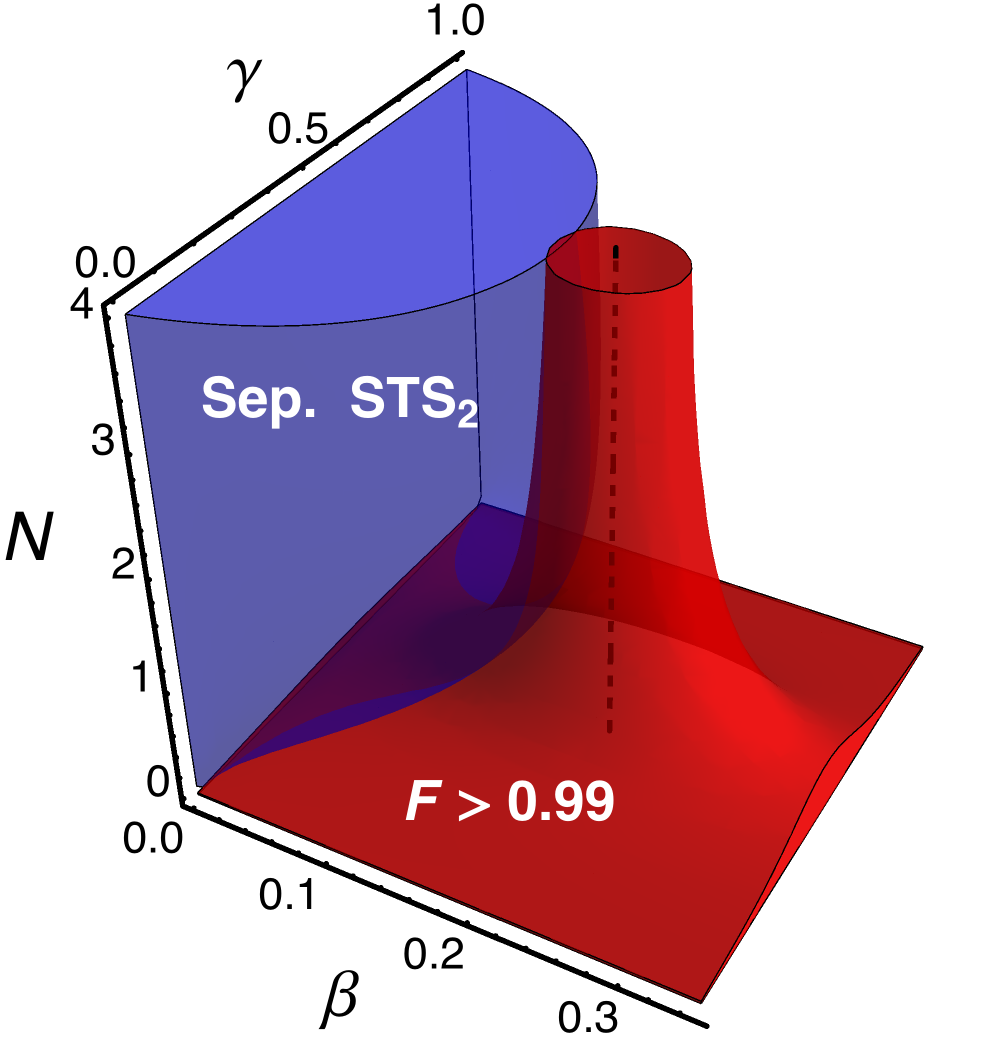}
}
\end{minipage}
\begin{minipage}{0.49\columnwidth}
\centerline{
\includegraphics[width=0.9\columnwidth]{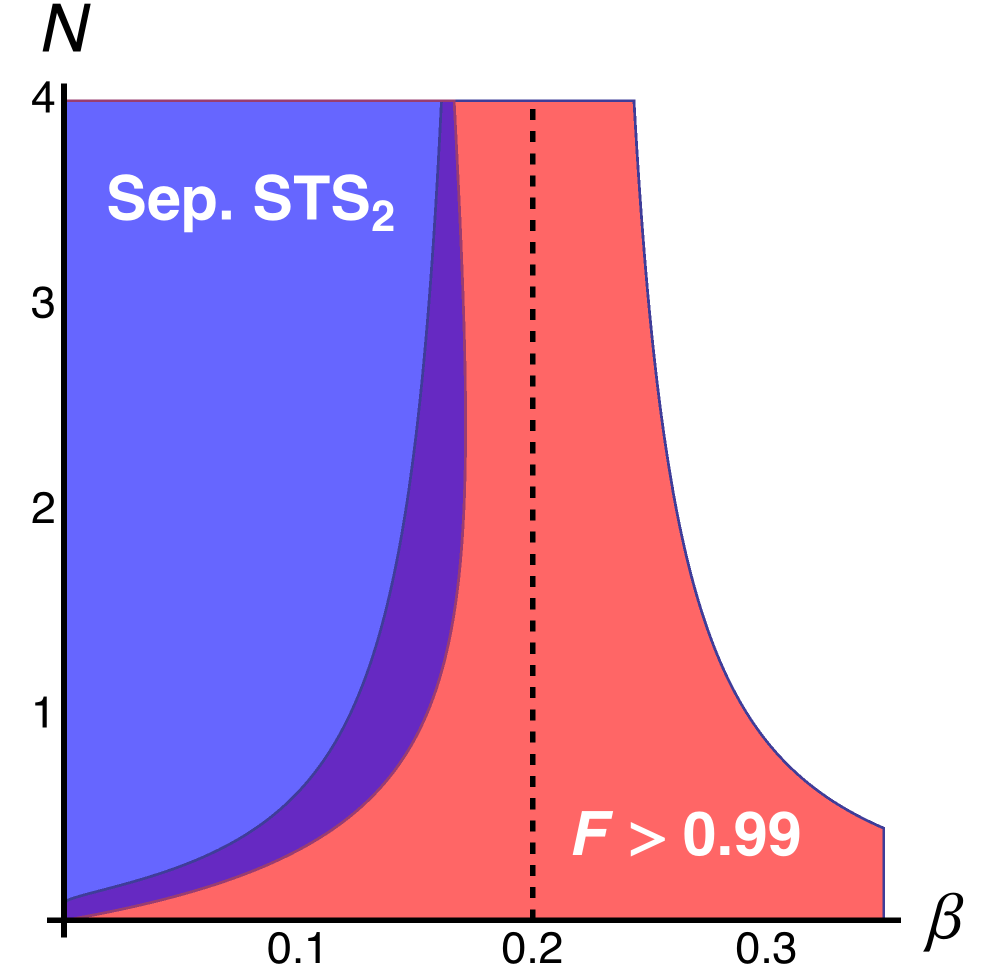}
}
\end{minipage}
\caption{\label{2mode}(Color online) (Left panel) Separability region of STS$_2$ (blue region)
in terms of the energy parameters $N$, $\beta$ and $\gamma$ on the left together with
the volume of states having $F_{N \beta \gamma}> 0.99 $ (red region) to a set
of entangled target STS$_2$ (black-dashed line) having the same energy $N$ 
with $\beta=0.2$ and $\gamma= 0.5$.
(Right panel) Section of the left panel plot in correspondence of $\gamma=0.5$. } 
\end{figure}
In the left panel of Fig.~\ref{2mode} we show the separability region in
terms of the three parameters $N$, $\beta$ and $\gamma$ and the
volume of states having $F_{N \beta \gamma} > 0.99$ with a set of
entangled target state with $\beta = 0.2$ and $\gamma= 0.5$. In order to
emphasize how the overlap is considerably large in the right panel we
have plotted a projection on the plane where it is maximized. The region
of separability is crossed by significant fraction of states over all the
energy range, thus making fidelity of a little use to asses entanglement
in these kind of systems though a severe constraint on the energy of the
two states has been provided.  
\par
\begin{figure}[h!]
\centering
\begin{minipage}{0.49\columnwidth}
\centerline{
\includegraphics[width=0.9\columnwidth]{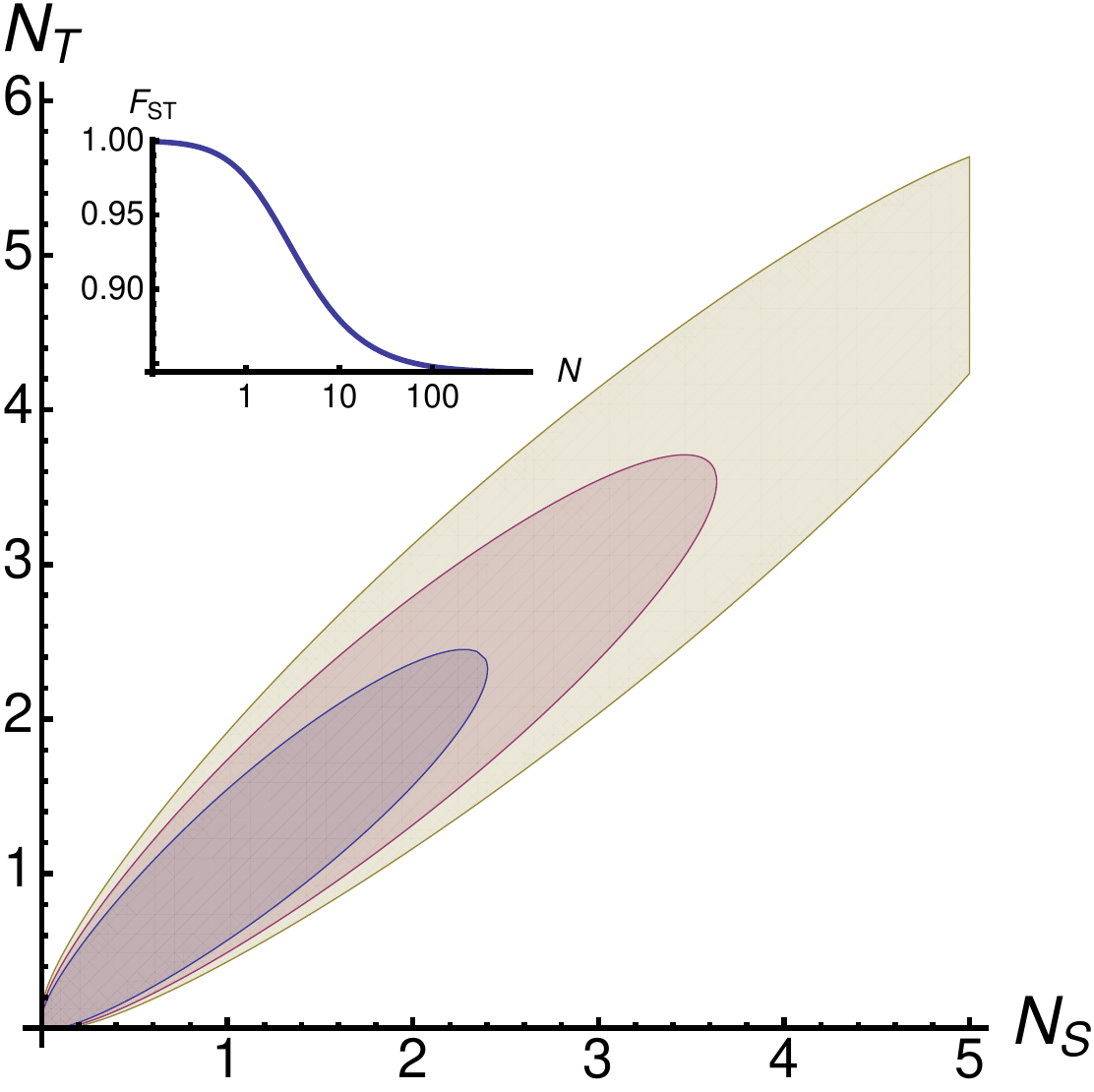}
}
\end{minipage}
\begin{minipage}{0.49\columnwidth}
\centerline{
\includegraphics[width=0.94\columnwidth]{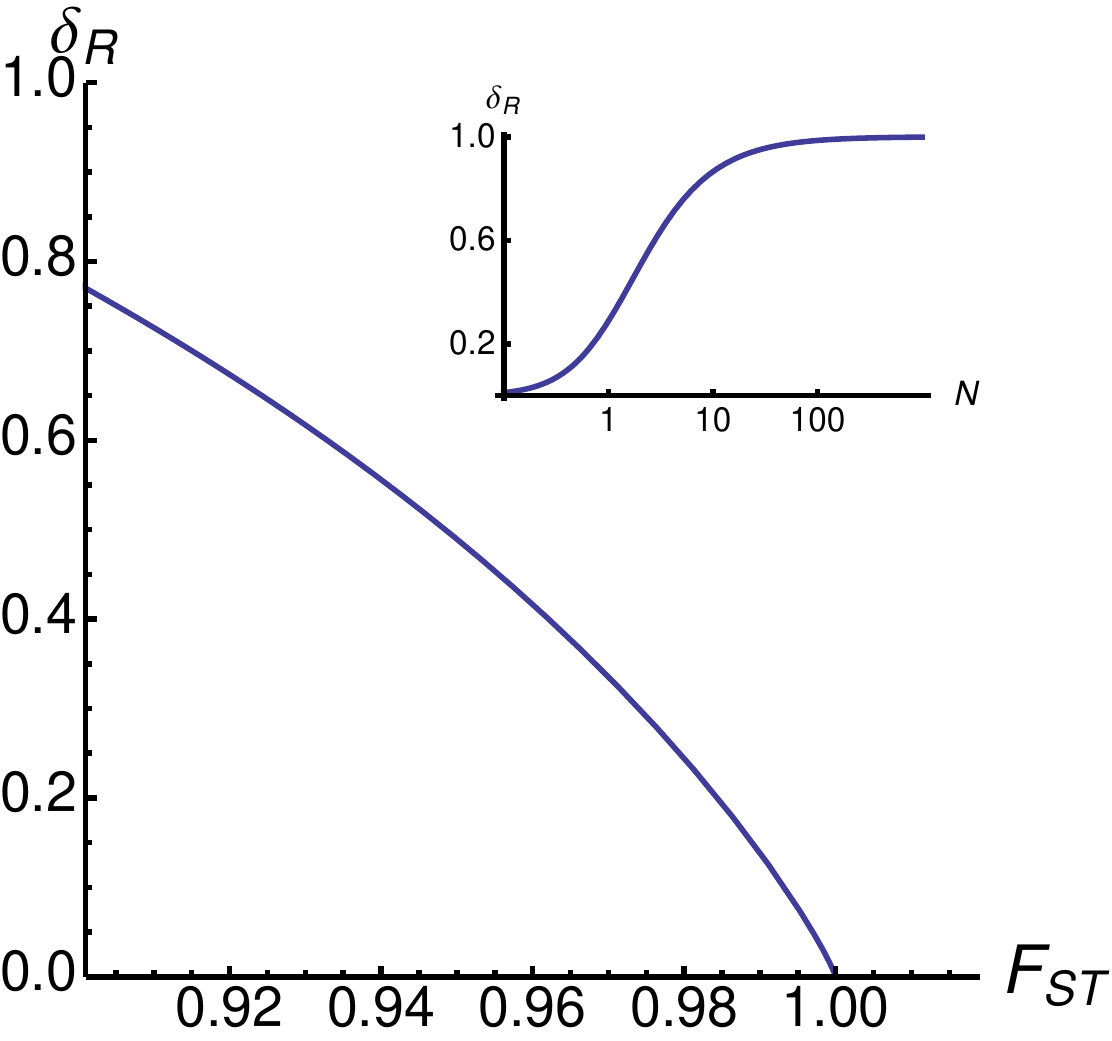}
}
\end{minipage}
\caption{\label{f:pnes} (Color online) (Left panel) Regions of states having fidelity larger
than 0.94, 0.92, 0.9 between a TWB and a PSSV in yellow,
red and blue, respectively; in the inset the logarithmic plot of fidelity
$F_{ST}$ in function of the energy, with $N=N_S=N_T$, which
reaches the value of $27/32$ in the limit $N \to \infty$. (Right panel)
Non-Gaussianity $\delta_R$ of PSSV as a function of the $F_{ST}$
to a TWB with same energy $N$; in the inset  the logarithmic plot
of  $\delta_R$ in function of $N$.} 
\end{figure}

As a final example, let us consider the set of photon-number 
entangled states (PNES), i.e. two-mode states
of the form~\cite{all1,all2}
$$
|\psi\rangle\rangle = \sum_n \psi_n |n,n\rangle\rangle\,,
$$
where $|n,n\rangle\rangle\equiv |n\rangle \otimes|n\rangle$. In
particular, we focus attention on two specific classes of PNES:
the Gaussian two-mode squeezed vacuum states (TWB)
$|\psi_T\rangle\rangle=S_2(r)|0\rangle\rangle$ and the non-Gaussian set
of states resulting from the process of {\em photon
subtraction}~\cite{wel00,mil02,ips03,ips04,ips04:b,ips04:c,kim05,wen04,par07} 
applied to $|\psi_T\rangle\rangle$, i.e. $|\psi_S\rangle\rangle\propto
\hat{a}\otimes \hat{b} |\psi_T\rangle\rangle$ (PSSV),
where $\hat{a}$ and $\hat{b}$ are the annihilation field operators.
In terms of the parameter $y=\tanh r$ we have:
\begin{align}
\psi_n^T = \sqrt{1-y^2}\,y^n 
\qquad \mbox{and} \qquad
\psi_n^S = 
\sqrt{\frac{(1 - y^2)^3}{1 + y^2}}\,(1 + n)\, y^n\,, 
\end{align}
such that the average numbers of photons are given by:
\begin{equation}
N_T = \frac{y^2}{1-y^2} \qquad \mbox{and} \qquad
N_S= \frac{2 y^2 (y^2+2)}{1-y^4}\,.
\end{equation}
In the left panel of Fig. \ref{f:pnes} we show some region plots 
of the fidelity  between a generic TWB and a generic PSSV
$$F_{ST} = \left|\langle\langle\psi_S|\psi_T\rangle\rangle\right|^2 = 
\left(\sum_n \psi_n^T\,\psi_n^S\right)^2\,,$$
as a function of their average number of photons. As it is apparent
from the plot, large values of fidelity, e.g.  $F_{ST}>0.9$, are compatible
with a relatively large range of energies, corresponding to considerably 
different physical properties (see below). Notice that for $N_T=N_S\equiv N$ we have 
$F_{ST}>27/32 \approx 0.84 $ $\forall N$: the inset shows the behaviour of $F_{ST}$ 
as a function of $N$ 
\par
A striking example of a property which cannot be assessed using 
fidelity is obtained by considering the non-Gaussianity of 
PSSV. For pure states the non-Gaussian character (quantum negentropy) of a CV
states may be quantified by the Von-Neumann entropy of its 
reference Gaussian state, i.e. a Gaussian state with the same
CM~\cite{nge,ngl,geong}.
For PNES the non-Gaussianity $\delta[\psi]$ reduces to
\begin{align}
\delta[\psi] = 2 
\left [ \left(d_{-}+\frac12\right) \log \left(d_{-}+\frac12\right)
- \left(d_{-}-\frac12\right) \log \left(d_{-}-\frac12\right) \right ]\,,
\end{align}
where  $d_-=\sqrt{(N+\frac12)^2-\big [\sum_n\, (1+n)\, \psi_n\, \psi_{1+n}\big ]^2}$.
The non-Gaussianity of PSSV is an increasing function of the energy.
In the right panel of Fig.~\ref{f:pnes} we show the non-Gaussianity
$\delta_R[\psi_S]$ of PSSV, renormalized to its asymptotic value 
(in order to have $0\leq\delta_R[\psi]\leq 1$)
as a function of the fidelity $F_{ST}$ between
the PSSV and a TWB with the same energy. As it is apparent from
the plot, very large values of fidelity to a Gaussian states are
compatible with very large values of non-Gaussianity.
The inset shows the behaviour of $\delta_R[\psi_S]$ as a function of
$N$.
\par
From our analysis, we conclude that also for two-mode states, fidelity should be
used with caution in order to assess quantum properties and that this 
is is true also when one imposes additional constraints on the
energy or the squeezing fraction of the states. Notice that also in 
the case of two modes, full tomography~\cite{FullCM,FullCM:b,bla12}
is imposing a suitable set of constraints to make fidelity a meaningful
figure of merit to summarize the overall quality of the reconstruction. 
\section{Conclusion}
\label{s:out}
In this paper we have presented several examples of single- and
two-mode CV states showing that being close in
the Hilbert space is by far not equivalent to share the same
physical properties, e.g. quantum resources. In addition, we have
shown that the phenomenon persists also when one imposes additional
constraints on the energy or the squeezing of the states, thus generally
questioning the use of fidelity to assess properties of CV systems.
Overall, our results suggest to use fidelity only in
conjunction with a tomographic set of additional constraints. 
\section*{Acknowledgments}
This work has been supported by MIUR through the FIRB project 
``LiCHIS'' Nr.~RBFR10YQ3H.

\end{document}